\documentclass[11pt,a4paper]{article}
\usepackage[hyperref]{eacl2021}
\usepackage{times}
\usepackage{latexsym}

\usepackage{graphicx}

\usepackage{comment}
\newcommand{\ligaikuma}{\textsc{Lig-Aikuma}}

\usepackage{microtype}

\aclfinalcopy 
\def\aclpaperid{15} 

\title{Fast Development of ASR in African Languages using Self Supervised Speech Representation Learning}

\author{
        Jama Hussein Mohamud%
        \thanks{  \quad African Masters of Machine Intelligence, Rwanda} \quad 
        Lloyd Acquaye Thompson\footnotemark[1] \quad Aissatou Ndoye\footnotemark[1] \quad \and 
        Laurent ~Besacier\thanks{   \quad Naver Labs, Europe}\\ \\
\{jmohamud, lthompson, andoye\}@aimsammi.org,\and laurent.besacier@naverlabs.com 
}

\date{}

\begin{document}

    \maketitle
    
\begin{abstract}
This paper is an informal collaboration launched during the African Master of Machine Intelligence (AMMI) program in June 2020. After a series of lectures and labs on speech data collection using mobile applications and on self-supervised representation learning from speech, a small group of students and the lecturer continued working on automatic speech recognition (ASR) projects for three languages: Wolof, Ga, and Somali. This paper describes the data collection process and ASR systems developed with a small amount (1h) of transcribed speech as training data. In these low resource conditions, pre-training a model on a large amount of raw speech is fundamental for developing an efficient ASR system. 
\end{abstract}

\section{Introduction}

Self-supervised learning using extensive unlabeled data has been explored with auspicious results in image processing ~\cite{norouzi}, and natural language processing  ~\cite{bert}. 
Pioneering works investigated self-supervised representation learning from speech ~\cite{Baevski, kawakami2020learning, chung2020generative}. They were successful in improving the performance on downstream tasks in speech recognition and low resource scenarios. 
In this paper, we investigate the possibility to leverage unlabeled speech for end-to-end automatic speech recognition (ASR) in an extremely low resource setting: only 1 hour of transcribed speech (quickly collected with a mobile application ~\cite{besacier:hal-02264418}) is available as supervised data to train an ASR system.

More precisely, we evaluate Contrastive Predictive Coding (CPC) ~\cite{Baevski} as an unsupervised speech representation method and apply it to ASR in 3 African languages: Wolof, Ga, and Somali.

\section{Related work}

Most deep learning methods highly rely on large quantities of labeled training data. Particularly, current acoustic models require thousands of hours of transcribed speech to get good performance, and this is a requirement lacked by the majority of the nearly 7,000 languages spoken worldwide ~\cite{lewis2009ethnologue}. To overcome this, self-supervised learning has been recently proposed as an interesting alternative for data representation learning while requiring less or no annotated data. Such learnt representations have been very successful in computer vision ~\cite{bachman2019amdim,chen2020simple,he2020momentum,dataefficient,misra2019selfsupervised} as well as in natural language processing ~\cite{devlin2019bert,peters2018deep, radford2018improving}. 

Self-supervised learning from speech consists of resolving pseudo-tasks not requiring human annotations as a pre-training to the real tasks to solve. These pseudo-tasks target predicting the next samples or solving ordering problems. Autoregressive predictive coding (APC) ~\cite{DBLP:journals/corr/abs-1904-03240, chung2020improved} considers the sequential structure of speech and predicts information about a future frame. An easier learning objective is introduced in  Contrastive Predictive Coding (CPC) which consists of distinguishing a true future audio frame from negatives ~\cite{Baevski, Schneider2019,morgane}.
~\citet{chung2020generative} shows that such representations are useful to improve several speech tasks while ~\citet{kawakami2020learning} extends those works by looking at the representations' robustness to domain and language shifts.
In the same vein, ~\citet{riviere2020unsupervised} compares self-supervised and supervised pre-training for ASR and shows that CPC pre-training extracts feature that transfer well to other languages, being on par or even outperforming supervised pre-training.
Another promising way is to use speech enhancement as a task for feature representation learning ~\cite{ravanelli2020multitask,ddsp}. 
Finally, several self-supervised tasks can be jointly tackled to discover better speech representations ~\cite{pascual2019learning}.

\section{Fast Data Collection using a Mobile App}

\subsection{Data collection}

Using applications on mobile devices, it is now possible to collect audio and video recordings from a large number of speakers with lower supervision from the researcher. Apps lower the pressure of defining the best sampling selection process, which speakers, and what data exactly to collect. Moreover, additional meta-information can be collected automatically from mobile devices (geographic coordinates, movement patterns, images, time codes). For instance, images (photos taken by potentially hundred of users) can be used to enrich lexical databases or, conversely, these images can be used to elicit speech.

In this work, speech recordings were elicitated from text using {\ligaikuma} ~\cite{lig-aikuma-sltu16,besacier:hal-02264418} which is  an improved version of the Android application \textit{Aikuma} initially developed by Steven Bird and colleagues ~\cite{bird_al_2014}. Features were added to the app to facilitate the collection of speech data. The result is a free Android app running on various mobile phones and tablets.  It proposes a range of different speech collection modes (recording, respeaking, translation, and elicitation) and offers the possibility to share recordings between users.
It can be downloaded from a dedicated website.\footnote{\url{http://lig-aikuma.imag.fr}} 

The data collection took place in June 2020, during the African Master of Machine Intelligence.\footnote{\url{https://aimsammi.org}} A speech recording project was conducted using  {\ligaikuma}. Each student (or the duo of students) recorded 2 hours of speech in their native language using the mobile app. This resulted in 30 repositories covering 19 African languages in total.\footnote{\url{https://github.com/besacier/AMMIcourse/tree/master/STUDENTS-RETURN}}

\subsection{The Three Languages}

The \textbf{Wolof} language is from the family of Niger-Congo, which belongs to the Atlantic branch. It is spoken in Senegal, Gambia, Mauritania and is the native language of the ethnic group of the Wolof people. Roughly ten million people speak Wolof. It has 29 consonant phonemes and 9 vowel phonemes. 

Approximately two million natives in Accra speak the \textbf{Ga}-Adangme language. It is part of the Kwa branch of the Niger-Congo language family. The Ga language has 31 consonant phonemes, 7 oral vowels, 5 nasal vowels with different vowel lengths. The origin of the transcribed text was extracted from disparate public sources through online and hard copy materials. The corpus includes Ga Literature, religious content from the Jehovah’s Witnesses website, poems, and songs composed in Ga. 

\textbf{Somali} is an Afroasiatic language belonging to the Cushitic branch. It is an official language of Somalia and Somaliland, a national language in Djibouti, and a working language in the Somali Region of Ethiopia and North Eastern Kenya. It is written officially with the Latin alphabet and it has 21 consonants and 5 vowel phonemes. The Somali text corpus is a collection of poems obtained from (Hadraawi).\footnote{\url{http://www.farshaxan.com/Gabayaa/hadraawi.html}}

In this paper, we focused on the three African language datasets (1h or 2h each) collected by the authors. The total utterances for all the entire datasets collected accumulate to 8359 as depicted in table \ref{2hr_data}. We split the data into train, dev, and test for each language and train our ASR systems on the train part and evaluate them on the test part.

We also benefited from a larger dataset (20h) in Wolof collected during a previous project by ~\citet{gauthier2016collecting}.\footnote{\url{https://github.com/besacier/ALFFA_PUBLIC/tree/master/ASR/WOLOF}} 

\begin{table}[!h]
\centering
\begin{tabular}{|l|l|l|l|l|l|l|}
\hline
\hline
Lang             & Train     & Dev     & Test    & \textbf{Total}\\  
\hline
Wolof            & 736       & 316     & 951     & \textbf{2003}            \\
\hline 
Ga               & 850       & 379     & 1402    & \textbf{2631}            \\
\hline
Somali           & 1241      & 725     & 1759    & \textbf{3725}            \\ 
\hline
\hline
\textbf{Total}   & \textbf{2827}       & \textbf{1420}    & \textbf{4112}    & 8359            \\
\hline
\end{tabular}
\caption{Number of Utterances for each small dataset collected using \ligaikuma}
\label{2hr_data}
\end{table}

\section{Self Supervised Representation Learning with Constrastive Predictive Coding (CPC)}

\begin{figure}[h]
    \centering
    \includegraphics[width=0.4\textwidth]{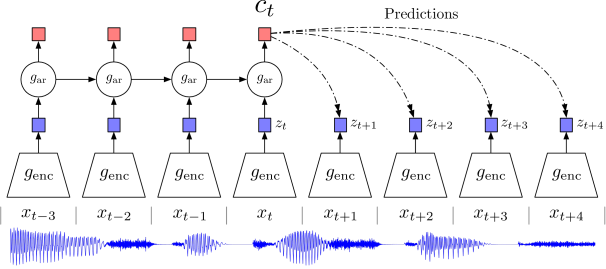}
    \caption{CPC Architecture: figure adopted from \citet{oord2019representation}. The non linear encoder $g_{enc}$ delineates the input sequence of observations $x_t$ to latent representations $z_t = g_{enc}(x_t)$. The auto regressive model $g_{ar}$ then condenses all the $z_{<=t}$ in the latent space to obtain a context latent representation $c_t = g_{ar}(z_{<=t})$, and $x_{t+k}$ depicts future observations whiles $x_{t-k}$ represents observed latent.}
    \label{CPC_architecture}
\end{figure}

CPC learns self-supervised features by predicting future latent features using powerful autoregressive models. We followed the approach of ~\citet{riviere2020unsupervised} and used the self-supervised pre-training model introduced in ~\citet{Schneider2019} (\textit{wav2vec}) which is based on contrastive predictive coding (see figure     \ref{CPC_architecture}).\footnote{Working on the recently introduced wav2vec2.0 ~\citet{baevski2020wav2vec} model is left for future work.}

The model uses (1) an encoder network that converts the audio signal into a latent representation, and (2) a context network that aggregates multiple time steps to build contextualized representations. 
The full model (encoder+context) is trained end-to-end to distinguish a sample that is k steps in the future from negative samples uniformly chosen from the same audio sequence. A contrastive loss is minimized for each step $k=1,...,K$ and the overall loss is summed over different step sizes (more details in ~\citet{Schneider2019}). 
The CPC architecture in Fig. \ref{CPC_architecture} obtained from ~\citet{oord2019representation} explains the flow of the CPC model in detail and, hence, we will urge readers to explore the original paper. 

 We evaluated the CPC model initially trained on a large English dataset by using it as a feature extractor for ASR using our three (1-2hr)  African datasets. We similarly trained a new CPC model on 20hr Wolof data and tested how it can transfer to ASR using the same (1-2hr) African datasets.

\section{Experiments and Results}

\begin{table*}[h]
\centering

\begin{tabular}{|c|c|c|c|c|c|}
\hline
\textbf{Model}        & \textbf{Pre-train} & \textbf{Frozen} & \textbf{wol}   & \textbf{so}  & \textbf{ga} \\
\hline
Linear/MFCCs  & No   & N/A    & 0.66 & 0.55 & 0.64 \\
\hline
CPC-English  & LL-60K   & Yes    & \textbf{0.32} & \textbf{0.26} & \textbf{0.34} \\
\hline
CPC-Wolof    & WOL-20h   & Yes    & 0.43 & 0.51 & 0.44  \\
\hline
CPC-Finetuned-Wolof    & LL-60K + WOL-20h   & Yes    & 0.39 &  0.48   &  0.41  \\
\hline
\end{tabular}
\caption{Transfer of self supervised pre-trained  models to ASR (measured as Phone Error Rate - PER on test sets) in 3 languages (wol:Wolof, so:Somali, ga:Ga). Linear/MFCCs: baseline using MFCC features. CPC English: pretrained model on 60k hours of English speech (from \cite{DBLP:journals/corr/abs-1912-07875}). CPC Wolof: pretrained model on 20 hours of Wolof data. CPC-Finetuned-Wolof: CPC English fine-tuned on 20h of Wolof data. For ASR we train a linear classifier on top of the frozen features using 1h for each language.}
\label{table:1}
\end{table*}

\begin{table*}[h]
\centering

\begin{tabular}{|c|c|c|c|c|} 
\hline
\textbf{Model}        & \textbf{Pre-train} & \textbf{Frozen} & \textbf{Wolof transcribed data} & \textbf{wol}  \\ 
\hline
CPC-English  & LL-60K & Yes   & 1h    & 0.32 \\ 
\hline
CPC-English  & LL-60K & No   & 1h    & 0.36 \\ 
\hline
CPC-English  & LL-60K & No  & 20h    & \textbf{0.28} \\ 
\hline
\end{tabular}
\caption{PER on Wolof ASR (test set) using 1h of transcribed speech / 20h of transcribed speech.}
\label{table:2}
\end{table*}

This section describes the approach used in the experiments and their outcomes which is documented in table \ref{table:1} and \ref{table:2} together with some useful observations made from those figures.

\subsection{Experiments}
Table \ref{table:1} reports Phone Error Rate (PER) for the four ASR models with Linear/MFCC as the baseline. MFCC (Mel Frequency Cepstral Coefficients) is used as a standard acoustic feature for speech recognition and speaker identification.
The model \textit{CPC-English} was pre-trained on a large English dataset (Libri-Light) with 60k hours of raw speech data \cite{DBLP:journals/corr/abs-1912-07875}. The \textit{CPC-Wolof} model is trained on the Wolof dataset of 20 hours as introduced earlier. In the \textit{CPC-Finetuned-Wolof} model, we initialized the CPC model's weight with that of \textit{CPC-English} model and then fine-tuned its parameters using the 20h Wolof corpus.

For ASR supervised training, we froze the model after the pre-training and simply learned a linear classifier for the targeted language using the smaller transcribed subsets (i.e. 1hr each) of the 3 languages. 
As explained in \cite{riviere2020unsupervised}, the linear classification is performed from the concatenation of 8 consecutive vector representations to match the average size of a phoneme. We then use the CTC loss \cite{10.1145/1143844.1143891} between our model predictions and the non-aligned phoneme transcriptions.
All the CPC backbone models were trained for 200 epochs and all the finetuning ASR experiments were done with 30 epochs.

\subsection{Results}
Our table shows that the best results (lowest PER) are obtained with the pre-trained model \textit{CPC-English} (60K), which confirms that a CPC model trained on a large data set can transfer well cross-lingually. On the other hand, training a CPC model on 20h of Wolof speech (\textit{CPC-Wolof}) is better than an MFCC baseline but worse than \textit{CPC-English} (60K) pre-training.
One explanation is probably that \textit{CPC-English} has been trained with more speaker diversity than \textit{CPC-Wolof}.
Surprisingly, adapting our CPC model to Wolof through fine tuning (\textit{CPC-Finetuned-Wolof}) does not improve the results either on Wolof ASR task.

We also investigated how using sizeable data for finetuning improves performance. 
Table \ref{table:2} reports PER outcomes for experiments done focusing on data sizes for the Wolof language. All attempts used the CPC model pre-trained on the very large English dataset (60k-hrs). We froze the parameters and used a 1hr Wolof transcribed data to obtain PER of 0.32 (already reported in table \ref{table:1}). 
Then, we used the \textit{CPC-English} model, added the linear classifier for phoneme recognition and then fine-tuned the full architecture (backbone model + linear classifier) using either 1h (2d line) in Wolof or 20h (3d line) in Wolof.
 
We can observe from the results that using a bigger amount of data (i.e. 20hrs) for finetuning does improve the PER (which reached its best PER=0.28). This improvement comes from the bigger data since fine tuning the full architecture only with 1h of transcribed speech did actually degrade the results compared to a frozen CPC model (compare lines 1 and 2 of the same table).

We also showed that languages originating from the same family produced indistinguishable results when using Wolof as the backbone. Wolof and Ga belong to the Niger-Congo Language family hence obtaining 0.43 and 0.44 respectively for PER. It seems that a model pre-trained on Wolof  could transfer well to other geographical nearby languages, but this claim would have to be confirmed with a backbone model trained on more Wolof data.

\section{Conclusion}
We presented a complete ASR development project for three languages from Africa: Wolof, Ga, and Somali. The project included data collection (using a custom mobile app), leveraging pre-trained models in a different language (English), and training phone recognition systems with a small amount of supervised data.

While cross-lingual transfer learning has shown a lot of success in speech and natural language processing, our work shows similar success in a couple of African languages. Our approach also outperforms the MFCC baseline, which indicates that the learned representation by CPC can build ASR systems even when we have limited (speech-text) data. 

As future works, the newly introduced wav2vec2.0 model \cite{baevski2020wav2vec} seems like a promising direction for further experiments on self-supervised representation learning for ASR.


\bibliography{anthology,eacl2021}
\bibliographystyle{acl_natbib}

\end{document}